\documentclass[10pt,journal]{IEEEtran}
\usepackage{amsmath, amssymb, bbm, mathtools, amsthm, bm}
\usepackage{graphicx}
\usepackage{subfigure}
\usepackage{verbatim}
\usepackage{soul}
\usepackage[]{footmisc}
\usepackage{enumitem}
\usepackage{tikz}
\usepackage[]{footmisc}

\usetikzlibrary{shapes,arrows}
\usepackage{xpatch}
\usepackage{hhline}
\usepackage{multirow}

\usepackage{algorithm,algorithmic}

\usepackage{cite}

\usepackage{cleveref}
\usepackage{tcolorbox}
\usepackage{diagbox}

\usetikzlibrary{arrows.meta}
\usetikzlibrary{positioning,fit,backgrounds}
  
  \tikzset{%
  >={Latex[width=2mm,length=2mm]},
            base/.style = {rectangle, rounded corners, draw=black,
                           minimum width=0.5cm, minimum height=1cm,
                           text centered, font=\sffamily},
  activityStarts/.style = {base, fill=black!30},
       startstop/.style = {base, fill=orange!25},
    activityRuns/.style = {base, fill=green!25},
         process/.style = {base, minimum width=1cm, fill=orange!25,
                           font=\sffamily},
            a/.style = {rectangle, rounded corners, draw=black, fill =green!30,
                           minimum width=1cm, minimum height=1cm,
                           text centered, font=\sffamily},
}

\usepackage{fancyhdr}
\begin{document}
% \maketitle
\title{Meta-model Neural Process for Probabilistic Power Flow under Varying N-1 System Topologies }
 
\author{Sel Ly, Kapil Chauhan, Anshuman Singh, and Hung Dinh Nguyen$^\star$
\thanks{$^\star$Corresponding Author. Authors are with School of EEE, NTU, Singapore, \textit{\{sel.ly, anshuman004, hunghtd\}@ntu.edu.sg}. Kapil Chauhan is with MNNIT Allahabad, India \textit{kapilchauhan@mnnit.ac.in}. This research is supported by the Agency for Science, Technology and Research (A*STAR), Singapore under its project M23M6c0114.}
%\\ Nanyang Technological University, Singapore \\
%\{\textit{sel.ly, kapil.chauhan, ehbgooi, hunghtd\}@ntu.edu.sg} 
%\\ $^\star$Corresponding Author}%
}
%\markboth{Published in Conference proceedings of IEEE PESGM 2023. DOI: 10.1109/PESGM52003.2023.10252798}%
 %{Shell \MakeLowercase{\textit{et al.}}}%
 
\IEEEaftertitletext{\vspace{-1\baselineskip}}

\maketitle

% Apply the fancy style to the first page, overriding the default plain style
\thispagestyle{fancy} 

\begin{abstract}
The probabilistic power flow (PPF) problem is essential to quantifying the distribution of the nodal voltages due to uncertain injections. The conventional PPF problem considers a fixed topology, and the solutions to such a PPF problem are associated with this topology. A change in the topology might alter the power flow patterns and thus require the PPF problem to be solved again. The previous PPF model and its solutions are no longer valid for the new topology. This practice incurs both inconvenience and computation burdens as more contingencies are foreseen due to high renewables and a large share of electric vehicles. This paper presents a novel topology-adaptive approach, based on the meta-model Neural Process (MMNP), for finding the solutions to PPF problems under varying N-1 topologies, particularly with one-line failures. 
% Once the MMNP model is trained, its accuracy can be maintained across different network configurations caused by N-1 contingencies. 
By leveraging context set-based topology representation and conditional distribution over function learning techniques, the proposed MMNP enhances the robustness of PPF models to topology variations, mitigating the need for retraining PPF models on a new configuration. Simulations on an IEEE 9-bus system and IEEE 118-bus system validate the model’s performance. The maximum $\%L_1$-relative error norm was observed as {1.11}\% and 0.77\% in 9-bus and 118-bus, respectively. This adaptive approach fills a critical gap in PPF methodology in an era of increasing grid volatility.

% The power supply networks undergo topology change due to multiple reasons including line outage, switching operation, fault isolation and reconfiguration. A topology change affects the accuracy of the conventional meta-models based probabilistic power flow (PPF) models which are typically trained on a fixed topology. 
\end{abstract}
% \begin{abstract}
% In real-time, power systems frequently undergo topology changes due to fault clearances, cyber-attacks, and scheduled maintenance, which significantly impacts the accuracy of conventional meta-models based probabilistic power flow. This paper proposes a novel topology-adaptive PPF approach that maintains model accuracy under dynamic network configurations. By leveraging graph-based neural networks and adaptive learning techniques, the proposed method enhances the resilience of PPF models to topology variations without the need for retraining on each new configuration. (See if to add one line about methodology) Simulations on an IEEE 9-bus system validate the model’s performance, showing a less than x\% error in probabilistic power flow predictions across various topological scenarios. This adaptive approach fills a critical gap in PPF methodology, enabling more robust, scalable, and responsive power system risk assessments in an era of increasing grid volatility.
% \end{abstract}


\vspace{-4mm}
\section{Introduction}
%\IEEEPARstart{D}{ue} 

Probabilistic Power Flow (PPF) analysis is a key technique in power systems, offering insights into network behaviour under uncertainty \cite{221262}. Unlike deterministic studies with fixed values for load and generation, PPF considers the variability from renewable energy integration and EV load. Further, PPF supports both planning and operational analysis. Specifically, in planning, it helps with the optimal placement of distributed energy resources and transmission line capacity assessment, while in operations, it aids in voltage stability evaluation, reliability analysis, and contingency preparation \cite{9516893}. 

% There are two primary approaches for solving PPF. The first, Monte Carlo (MC-PPF), is a conventional method that involves running multiple power flow simulations (N scenarios) to capture probabilistic outcomes across uncertain inputs like variable renewable generation and demand. 

The power flow behaviour under uncertain scenarios can be obtained through a Monte Carlo-based approach which involves generating a large number of samples for uncertain parameters followed by a load flow computation for each scenario. This approach provides distributions of key variables, such as bus voltages and line flows. However, its complexity increases linearly with the number of scenarios, Monte Carlo-based probabilistic power flow (MC-PPF) can be computationally demanding for large networks or high-resolution operational studies \cite{8447243}. 
% To mitigate this, researchers have developed 
An alternative approach is to utilize meta-models (or surrogate models) which can be trained on a smaller subset of scenarios.
% , M (where M $<$ N), model-based PPF (MM-PPF). This approach
By capturing relationships between uncertain inputs and power flow outputs, meta-model-based probabilistic power flow (MM-PPF) significantly reduces computational requirements while maintaining sufficient accuracy.
Meta-models applied to PPF are typically constructed using machine learning techniques such as polynomial chaos expansion \cite{9870032}, modified polynomial chaos expansion \cite{10252798}, Gaussian process regression \cite{pareek2021non,optimalcontrol}, and neural networks. These techniques have been found to be effective in approximating the complex, nonlinear relationships in power flow equations. Additionally, these models have shown promise in reducing computational requirements by orders of magnitude, enabling faster PPF studies for large networks. 
% For instance, a meta-model trained on M samples can provide predictions for voltage magnitude and power flows for N (N$>$M)  scenarios, which is highly advantageous for real-time applications. 

Despite their advantages, meta-model-based PPF approaches face a significant challenge: \textit{sensitivity to changes in network topology}. Since meta-models learn from a fixed set of scenarios based on a specific network configuration, they may lose accuracy when the topology changes. 
% Network topology changes frequently in operational contexts due to events such as fault clearances \cite{10192358}, system load balancing \cite{10129094}, maintenance activities \cite{9619978}, power loss minimization \cite{9521478}, cybersecurity incidents \cite{8118126}, etc. 
A network change in the power system may occur due to multiple reasons including line or equipment outage, switching operations for load balancing or maintenance \cite{9619978}, fault isolation, cyber-security incidents, etc. Further, the recent push for higher penetration of renewable energy sources may increase the switching operations due to their high variability. One approach for handling topology variation is to train a separate model for each topology. However, the number of possible power network configurations increases combinatorially, posing substantial computational challenges for learning, particularly when combined with load uncertainties \cite{gao2023physics}.
% When these changes occur, line configurations, bus connections, and load or generation patterns may alter, impacting the meta-model's accuracy in predicting power flow outcomes.

A simple way to reflect topology changes is by adding branch status (0 or 1) as a feature, but this ignores line outage effects on connected lines \cite{xiang2020probabilistic}. Another approach is to utilize the admittance matrix, but the transmission line parameters are not always accurately available. Authors in \cite{xiang2020probabilistic} proposed a deep neural network-based PPF model and utilized the voltage difference of each bus before and after contingency as an input feature. Another approach proposed in \cite{du2019achieving} uses diagonal elements of the admittance matrix as a training feature.

In this paper, we focus on topologies resulting from N-1 contingencies of line outages. However, we limit the situation when no bus is islanded due to any line outage. The trained model will be able to capture both power flow relationships and the topology. 
% For a new power injection variation (due to renewables, for example) in a learned N-1 topology, the trained model is able to produce the distribution of the nodal voltages, i.e., probabilistic power flow solution.
As per the best of authors' knowledge, existing literature largely considers the network topology to be static in PPF studies. This paper proposes a novel approach for PPF that is topology-adaptive using Neural Processes (NPs) \cite{garneloNP}, allowing the meta-model to maintain accuracy even as the network configuration changes. NPs are an advanced class of probabilistic models that integrate the advantages of Gaussian Processes (GPs) and Neural Networks (NNs) to address prediction uncertainties by meta-learning distributions over functions. This approach enables flexible data representation and enhances generalization. NPs can quickly adapt to new data, making them highly effective in scenarios that demand rapid adjustment \cite{brunzema2024neural}. For a comprehensive survey of the NP family, we refer to the paper \cite{jha2022neural}. % By incorporating flexibility for topology variations, this approach addresses a significant gap in current methods and enhances the robustness and applicability of PPF in real-world dynamic power systems.
The main contributions of this paper can be summarized as follows.
\begin{enumerate}
    \item Propose topology representation based on \textit{context set} for presenting different network topologies due to N-1 contingencies. Here, each \textit{context set} consists of a sample of input-output pairs encoded into a latent representation for each topology by using encoder networks. 
    
    %By encoding this pair into a latent representation using encoder networks, the meta-model can learn to capture the underlying dependence structure between inputs and outputs for each topology.  
    
    \item Develop a meta-model Neural Process for learning probabilistic power flow solutions under varying N-1 system topologies. 
    %This model combines the power of neural networks  and Gaussian Process to form a NN-based stochastic process for predictions and uncertainty quantification. 
    During inference, each target input is referred to as a query, and then the predictive distributions of target outputs are made by conditioning on both this query and the context set representation of the topology. 
    \item Partition ``similar'' N-1 contingency topologies into clusters; for each cluster, a meta-model Neural Process for PPF will be trained. This clustering practice helps improve the performance of the trained MMNP model within the respective N-1 topology cluster.
\end{enumerate}

The remaining paper is organized as follows. Section II develops the proposed NP-based PPF. Section III provides the results and discussion on the test cases. 

 \vspace{-0.5em}
 
\section{NPs for Probabilistic Power Flow} \label{sec:methods}
% We first briefly formulate the PPF problem. Next, we introduce NP background and describe \textit{context sets} used to represent different topologies.  The proposed meta-model NP  is then trained and applied for predicting all nodal voltage outputs considering changes in N-1 system topologies and uncertainties in load demands.

%\vspace{-0.3cm}
\subsection{Probabilistic power flow (PPF) problem } \label{sec:PPF_N_1}
% In this work, we aim to study PPF under system topology reconfiguration, i.e., obtaining the distribution of state variables (bus voltage and power flow) when there is the uncertainty of nodal power injection and varying system topology. 
In a deterministic power flow analysis for a $N$-bus system, when given a vector of load demands (input vector): $\boldsymbol{\xi}:=[\bm{P},\bm{Q}]=[P_1,\ldots,P_d, Q_1, \ldots, Q_d]$, we can determine a vector of  voltage states (output vector), $\boldsymbol{V}:=[V_1,\ldots,V_N]$ by  solving a set of non-linear power flow equations of the AC power flow problem, see \cite{HEIDARI2022107592,nguyen2018constructing}:   
%  \begin{equation}
% g(\boldsymbol{v},\boldsymbol{\xi},\boldsymbol{C})=0, \label{sec:pf}
%  \end{equation}
 %\cite{nguyen2018constructing}
\begin{align}
P_k + j Q_k & = V_k \sum_{l=1}^{N} Y_{kl} \,V_l\, \exp(-j \theta_{kl}) ,\quad k \in  \, \mathcal{L,G}, \label{eq:PF} 
\end{align}
where the notations are defined as follows:
\begin{itemize}
    \item $P_k/Q_k :$ active/reactive power injections at bus \textit{k}. 
\item $V_k, \theta_k :$ voltage magnitude and phase at bus \textit{k}.
\item $\theta_{kl} = \theta_k - \theta_l:$ phase difference between buses $k$ and $l$. 
\item $Y_{kl}$: element $(k,l)$ of admittance matrix Y. 
\item $ \mathcal{L}$ and $ \mathcal{G}$ are sets of all loads and generators in network.
\end{itemize}

In practice, the load demand $\bm{\xi}$ varies due to various factors such as weather conditions, time of day, and consumer behaviour. These fluctuations introduce uncertainties that can affect the output distributions such as voltages. Topology changes, such as the addition or removal of transmission lines, also affect the voltages and line flows.

%\vspace{-0.3in}
\vspace{-1em}

\subsection{Proposed NP-based PPF under varying system topology}

\subsubsection{Background of Neural Processes (NPs)}

\begin{figure}
    \centering    \includegraphics[width=0.96\linewidth]{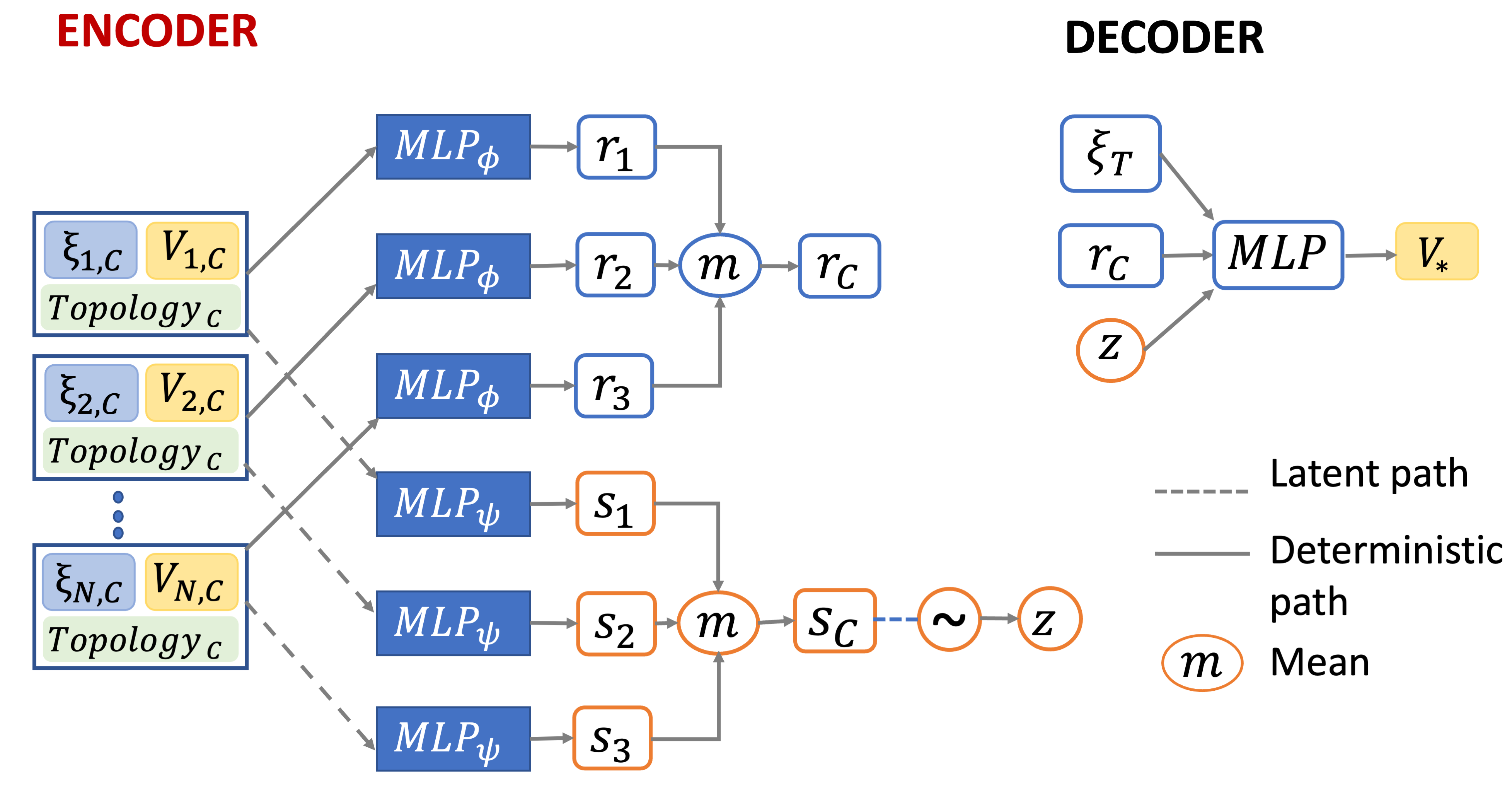}
    \caption{System architecture for the NP model.}
    \label{fig:np}
    \vspace{-2mm}
\end{figure}

%A Neural Process (NP) is an advanced probabilistic machine learning model proposed in \cite{garneloNP} that combines the flexibility of neural networks with the probabilistic framework of Gaussian processes (GPs) to represent a novel NN-based stochastic process. This model can be designed to handle tasks that require learning from a few examples, which is particularly useful for meta-learning and few-shot learning scenarios \cite{brunzema2024neural}.
%designed to improve the performance of a traditional Neural Process (NP) by incorporating attention mechanisms \cite{kimattentive} which is particularly used in transformer models, including Large Language Models (LLMs). The attention mechanisms allow the model to focus on relevant parts of the input data when generating each part of the output. 

%In particular, the standard NP was first p using neural networks to represent a stochastic process as an alternative to GP. 
In this paper, we propose NP  \cite{garneloNP} that learns to map a set of observed input-output, called the \textit{context set} $(\bm{\xi}_C, \bm{V}_C):= \{([\bm{p}_i, \bm{q_i}],\bm{v}_i)\}_{i \in C}$, to determine the conditional distribution over functions of a set of target outputs $(\bm{\xi}_T, \bm{V}_T):= \{([\bm{p}_i,\bm{q}_i], \bm{v}_i)\}_{i \in T}$, where $C$ denotes a set of observed \textit{context points} and $T$ represents a set of unobserved \textit{target points}. In the meta-learning algorithms, each context set $C$ represents each task to be learned, and the model can learn to adapt several related tasks. Similarly, we propose to use the context sets to represent topology structures and the tasks are to learn the PPF problem for these topologies. The predictive conditional distribution of the target output $\bm{V}_T$ can be formulated as:
\[
p(\bm{V}_T | \bm{\xi}_T, \bm{\xi}_C, \bm{V}_C) := \int p(\bm{V}_T | \bm{\xi}_T, r_C(z)) q(z | s_C) \, dz,
\]
where $Z$ is a global latent variable designed to capture the predictive uncertainty. The generative framework of the NPs is given step by step as follows, see Fig. \ref{fig:np}:

\begin{enumerate}
    \item[(i)] For the \textit{deterministic path}, we encode the latent representation for all input-output pairs in the context set $C$ using a multi-layer perceptron (MLP) encoder network:  $\bm{r}=Enc_{\bm{\phi}}(\bm{\xi}_C,\bm{V}_C)$.
    \item[(ii)] We aggregate vectors $\bm{r}$ (normally using mean) to obtain a single value $r_C$.
    \item[(iii)] For the \textit{latent path}, we apply another encoder network:  $\bm{s}=Enc_{\bm{\psi}}(\bm{\xi}_C,\bm{V}_C)$, and similarly  aggregate it into $s_C$. 
    \item[(iv)] The global latent variable Z is then parametrized by: $p(z|\bm{\xi}_C,\bm{V}_C)= p(z|\bm{s}_C)=N\big(\mu(s_C),\sigma^2(s_C)\big)$. 
    \item[(v)] Finally, we pass the tuple $(\bm{\xi}_T,r_C,z)$  through the MLP decoder network:
$\big(\mu(\bm{V}_T),\sigma(\bm{V}_T)\big)=Dec_{\bm{\theta}}\big(\bm{\xi}_T,r_C,z\big)$, to predict both the mean and standard deviation of the target output $\bm{V}_T$.
\end{enumerate}

% this latent variable is modelled as a Gaussian and  parameterized by $s_C = s(\bm{\xi}_C, \bm{V}_C)$ through a multi-layer perceptron (MLP) encoder network, see \cite{} for more detail. % $q(z|s_C) = N(z| \mu_z, 0.1+0.9\sigma(\omega_z) $

% In standard NPs, a deterministic representation $r_C^*$ aggregates information across the context set, typically by taking average, i.e. the mean-aggregation. For ANPs, however, we utilize \textit{cross-attention mechanism} instead, see Fig. \ref{fig:anp}. Note that in this paper, we use the notation $\bm{\xi}$ for the input (not $x$) and $\bm{V}$ for the output (not y). In the \textit{cross-attention mechanism}, each target $\bm{\xi}_T$ is referred to as a query that attends to the context set $(\bm{\xi}_C, \bm{V}_C)$, represented as $r_C^*:= r(\bm{\xi}_C, \bm{V}_C, \bm{\xi}_T)$, creating a query-specific representation. Finally, we pass the tuple $(\bm{\xi}_T,r^*_C,z)$  through the MLP decoder network to  predict both the mean and standard deviation of the target output $ \bm{V}_T$. For more detail about development and deployment of ANP models, we refer to \cite{}.
%This enhancement preserves the global structure of the process via the latent variable $z$ in the generative path while the deterministic path captures the local details. 

The model parameters (e.g. $\bm{\phi}, \bm{\psi}, \bm{\theta}$) are learned through variational inference, maximizing the evidence lower bound (ELBO):
\begin{align}
    \log p(\bm{V}_T | \bm{\xi}_T, \bm{\xi}_C, \bm{V}_C) & \geq \mathbb{E}_{q(z | s_T)}[\log p(\bm{V}_T | \bm{\xi}_T, r_C, z)] \nonumber \\
   & \quad - \text{KL}(q(z | s_T) \| q(z | s_C)),
\end{align}
where a variational distribution $ q(z|s_T)$ is used to approximate the intractable prior $ p(z|s_T) $. In addition, the Kullback–Leibler (KL) divergence is utilized for encouraging alignment of context and target sets within the same data-generating process. 

\subsubsection{Proposed Context Set-based topology representation}
Recall that we limit the situation when no bus is islanded due to N-1 contingencies to avoid infeasible load flow issues. Here, we propose a novel approach to learn voltage profiles under these varying system topologies using the proposed NPs. The detailed approach through three stages of data collection, model training and model prediction, is given as follows:

\begin{enumerate}
    \item[(i)]\textbf{Data Collection}: 
    %\begin{itemize}
       %\item \textbf{Context Sets}: 
        For each system topology, we collect a sample dataset consisting of input-output pairs. This sample data are then randomly split into \textit{context sets} $(\bm{\xi}_C, \bm{V}_C)$ and \textit{target sets} $(\bm{\xi}_T, \bm{V}_T)$. Normally, the context set is chosen as a subset of the \textit{target set}. 
        %where inputs are load demands, and outputs are the corresponding voltage profiles.
%\item \textbf{Multiple Topologies}: 
We gather \textit{context sets} and \textit{target sets} for various topologies forming a training dataset.
%ensuring a diverse and comprehensive dataset that captures different network configurations and their impacts on voltage profiles.
    %\end{itemize}
    \item[(ii)]  \textbf{Model Training}:
%\begin{itemize}
    %\item \textbf{Combined Dataset}: The collected context sets from different topologies are combined to form a comprehensive training dataset. This dataset is used to train the proposed ANP model.
%\item \textbf{Training Process}: 
During training, the proposed NP learns to map the input context points to the corresponding output target points in the training dataset above. 
%The cross-attention mechanism within the ANP allows the model to focus on relevant context points and target points, improving its ability to generalize across different topologies.
%end{itemize}
\item[(iii)] \textbf{Prediction and Evaluation:}
%\begin{itemize}
    %\item \textbf{Specific Topology and Target Set}: 
    After training, given a specific system topology and a new target set of inputs $\bm{\xi}_T^*$, we select a sample context set $(\bm{\xi}_C, \bm{V}_C)$ from the training dataset corresponding to this topology.
%\item \textbf{Combining Context and Target}: 
% The selected context set is then combined with the target inputs, and 
The trained NP takes this tuple $(\bm{\xi}_T^*, \bm{\xi}_C, \bm{V}_C)$ as the input to predict the unobserved target outputs of voltage profiles $\widehat{\bm{V}}_T^*$. %The cross-attention mechanism enables the ANP to effectively utilize the relevant context points and target points, ensuring accurate and reliable predictions.
Then, the prediction performance of the proposed model is evaluated using the $\%L_1$-relative error norm, and two well-known metrics: RMSE and MAE, defined as:
% \begin{align}
% \%\Vert \Delta \bm{V}_T^* \Vert_1:=\frac{\Vert \widehat{\bm{V}}_{T}^* - \bm{V}_{MCS} \Vert_1 }{\Vert\bm{V}_{MCS} \Vert_1}\times 100, \label{eq:L1-error}
% \end{align}
\begin{align}
\%\Vert \Delta \bm{V}_T^* \Vert_1 &:=\frac{\Vert \widehat{\bm{V}}_{T}^* - \bm{V}_{MCS} \Vert_1 }{\Vert\bm{V}_{MCS} \Vert_1}\times 100, \label{eq:L1-error} \\
    \text{RMSE}(\bm{V}_T^*) &= \frac{1}{N}\sum_{j=1}^N\sqrt{\sum_{i=1}^{n}\frac{(\widehat{V}_{j,i}^*-V_{j,i})^2}{n}}, \\ %\quad  RMSE(\bm{V}_T^*) = \frac{1}{N}\sum_{j=1}^N  \text{RMSE}(V_j), \\
     \text{MAE}(\bm{V}_T^*)&= \frac{1}{N}\sum_{j=1}^N\sum_{i=1}^{n}\frac{|\widehat{V}_{j,i}^*-V_{j,i}|}{n},
\end{align}
% \begin{align}
%     \text{MAE}(V_j) = \frac{1}{n}\sum_{i=1}^{n}|\widehat{V}_{j,i}^*-V_{j,i}|, \quad  \text{MAE}(\bm{V}_T^*) = \frac{1}{N}\sum_{j=1}^N  \text{MAE}(V_j). 
%\end{align}
where $\bm{V}_{T}^*$ and $\bm{V}_{MCS}$ represent the nodal voltage vectors predicted by the proposed method and generated by Monte Carlo simulation (MCS), respectively. 
%\end{itemize}
\end{enumerate}

% We argue that the effectiveness of this approach may lie in the cross-attention mechanism of the ANP. By allowing the model to dynamically focus on the most relevant context points for each query from target points, the ANP can capture the intricate dependencies and variations in voltage profiles due to changes in system topology. This results in a robust and flexible model capable of handling the complexities of modern power systems.

\vspace{-1em}
\section{Results and discussion} 
\label{eq:results}

\subsection{System Description and Data Generation}
The proposed method was first validated on the IEEE 9-Bus test system \cite{chow1982time} and later on IEEE 118-Bus network \cite{118_bus}. 
% The single-line diagram for this network is shown in Fig. \ref{fig:9bus-net}. 
The 9-Bus network consists of 3 generators at buses 1, 2, and 3. The load profiles were generated considering normal distribution with a mean and standard deviation of 0.9 p.u. and 0.05 p.u., respectively. Additionally, a correlation coefficient of 0.5 was considered between bus-demands. The network consists of a total of nine transmission lines, out of which six lines are feasible for N-1 contingency analysis. This results in seven distinct topologies, as outlined in Table I. Simulations were performed on a computer using MATLAB with the MATPOWER toolbox \cite{zimmerman2010matpower}.
% \begin{figure}[tbp]
%     \centering
%     \includegraphics[width=0.7\linewidth]{figs/9bus_network.png}
%     \vspace{-2mm}
%     \caption{IEEE 9-bus network.}
%     \vspace{-5mm}
%     \label{fig:9bus-net}
% \end{figure}

\begin{figure}[!tbp]
   \centering
\includegraphics[width=\columnwidth]{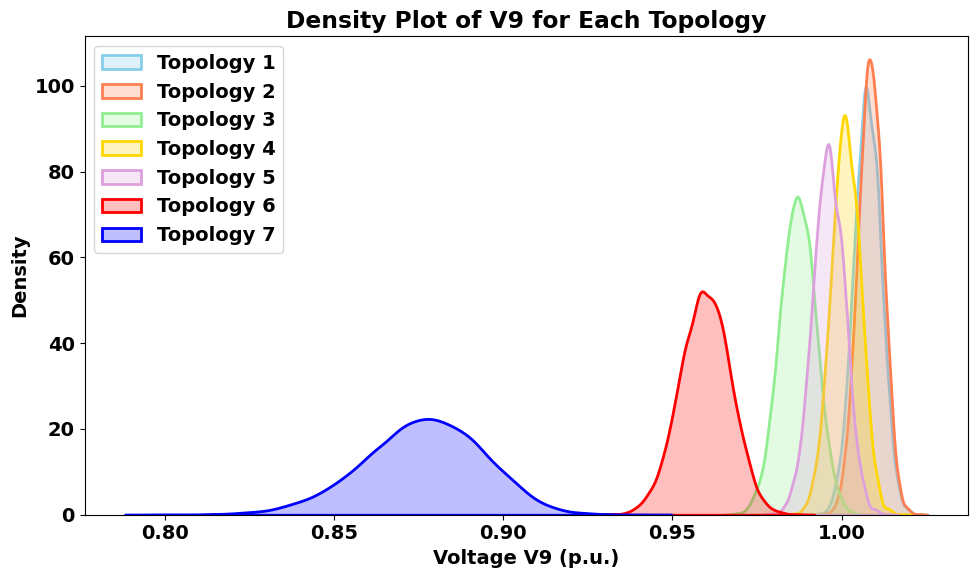}
   \vspace{-8mm}
   \caption{Comparison of distributions of voltage at bus 9 $(V_9)$ under seven different topologies of IEEE 9-bus network, given the same load distribution.}
   \label{fig:topo_Vdenst}
   \vspace{-2mm}
\end{figure}

\begin{figure}[!tbp]
   \centering
\includegraphics[width=.9\columnwidth]{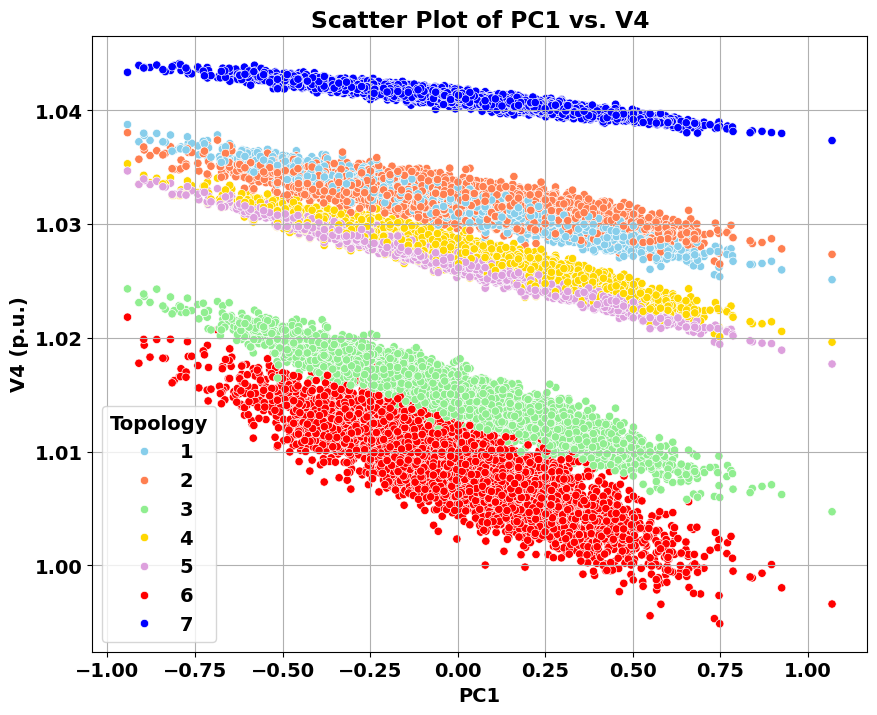}
   \vspace{-3mm}
   \caption{Comparison of voltage at bus 4 $(V_4)$ under seven different topologies of IEEE 9-bus, given the same loads. Note: PC1 is the main principal component of the vector of load inputs.}
   \label{fig:topo_Vpc}
   \vspace{-6mm}
\end{figure}

\begin{table}[tbp]
    \centering
    \vspace{-2mm}
    \caption{List of N-1 System Topologies.}
    \vspace{-2mm}
    \label{tab:topos}
    \begin{tabular}{c|c}
    \textbf{Topology} &  \textbf{Line Outage} \\
    \hline
        1 &  Base case (no line outage)\\
         2 & Line from bus 4 to bus 5 \\
           3 & Line from bus 5 to bus 6 \\
             4 & Line from bus 6 to bus 7 \\
               5 & Line from bus 7 to bus 8 \\
                 6 & Line from bus 8 to bus 9 \\
                   7 & Line from bus 9 to bus 4 \\
                   \hline 
    \end{tabular}
\end{table}
 % It is observed that for the same load scenario, the voltage profile of the network is influenced by the topology.
\vspace{-1em}
 Fig. \ref{fig:topo_Vdenst} presents the voltage density at bus 9 for various load scenarios under each topological configuration. As observed, each topology exhibits a distinct density curve, with the mean voltage ranging from {0.875 p.u. to 1.05 p.u.} It can also be observed that the voltage of bus 9 under Topology 7 is significantly lower than bus voltages under other topologies. 
 % This is due to the fact that under topology 7, the demand at bus 9 is fulfilled through line 7 (i.e. line connecting bus 8 to bus 9) whose resistance and reactances are significantly higher (0.032	and 0.161 p.u. respectively).
 Further, Fig. \ref{fig:topo_Vpc} displays a scatter plot between the main principal component of the same load inputs versus the bus 4 voltage under varying seven N-1 system topologies, where distinct clusters along the PC1 axis indicate the influence of topology on the voltage profile. This supports the argument that a conventional model trained for a certain topology is not accurate enough for another topology of the same network. Next, we show the performance of the proposed meta-model NP that can capture the changing topology through conditioning on \textit{context sets}. 

 %\vspace{-1em}
 
We design the proposed MMNP architecture using MLP encoder and decoder networks, each consisting of 4 layers with 32 hidden nodes per layer. Also, the dimension of the context point representation $\bm{r} $ and the sampled latent variable $Z$ are both set to 32. The model is trained with the Pytorch framework using a batch size of 4, a learning rate of 3e-4 and a fixed number of 1000 epochs. During training, we randomly selected 30 context points and 30 target points to construct each batch, thereby capturing the characteristics of each topology. \textbf{Also, it is crucial to normalize both the input $\bm{\xi}$ and the output $\bm{V}$ using a standard scaler \emph{separately for each topology} prior to training.}

%For each topology, we collected 1000 sample data points for the training and the remaining data of 4000 points was used for testing. 
% During the training, we randomly selected 30 number context points and 30 number target points to form each batch (capturing each topology). \textbf{It is very important to note that we have to normalize (standard scaler) for both the input $\bm{\xi}$ and the output $\bm{V}$ for each topology separately before  the training. } 

\vspace{-0.5cm}
\subsection{Accuracy Analysis } 
This section details out the clustering approach and the performance analysis for the proposed model. The proposed model is analyzed on two scenarios as follows:

\textbf{Case 1}: Training only one proposed meta-model NP for all seven N-1 topologies. 

\textbf{Case 2}: Training three proposed NPs for the corresponding three topology clusters as suggested by Fig. \ref{fig:topo_Vpc}. Specifically, \\
\textit{Cluster 1}: one proposed NP for Topology 2, 4, and 5.  \\
\textit{Cluster 2}: one proposed NP for Topology 3 and 6. \\
\textit{Cluster 3}: one proposed NP for Topology 7.

We adopt a clustering-based approach instead of training a single model on the entire dataset. Grouping similar topologies into clusters enhances model performance by capturing shared patterns and behaviours. Clustering enables the model to learn from smaller, homogeneous subspaces, significantly improving prediction accuracy, especially in large networks. It also captures topology-specific features, such as a generator bus's influence on voltages, which are harder to learn from the entire dataset. In this work, clusters are designed based on contingency severity, quantified by post-contingency bus voltages. For example, in Fig. \ref{fig:topo_Vpc}, topology 7, with significant voltage deviations, is treated as a separate cluster, while topologies 3 and 6, and 2, 4, and 5 are grouped into clusters. 

\begin{figure}[tbp]
\begin{center}
% \subfigure[ Topology 1.]{
% \includegraphics[scale=0.5]{figs/topo1_hist_error.png}
% \label{fig:topo1}}
\subfigure[ Topology 2 (Case 1). ]{
\includegraphics[height = 4.5cm,width=7cm]{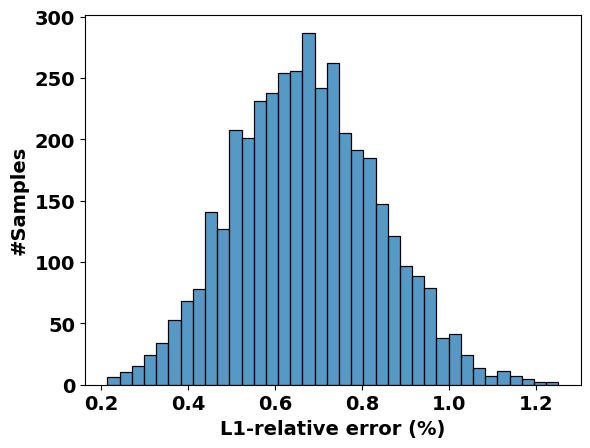}
\label{fig:topo2}}
\subfigure[ Topology 7 (Case 2). ]{
\includegraphics[height = 4.5cm,width=7cm]{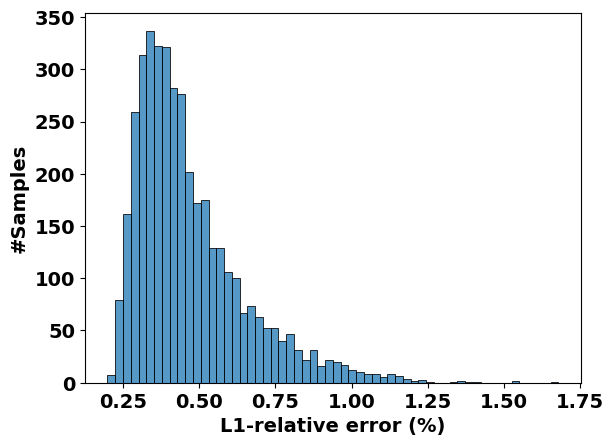}
\label{fig:topo3}}
\vspace{-2mm}
\caption{L1-relative error (\%) in predicting voltage magnitudes under different topologies of IEEE 9-bus network using the proposed NP models. }
\label{fig:diff_topo}
\end{center}
\vspace{-8mm}
%\vspace{-1.0em}
\end{figure}

The performance of the proposed meta-model is evaluated using the $\%L_1$-relative error norm defined in \eqref{eq:L1-error}. 
Fig. \ref{fig:diff_topo} shows histograms of the $L_1$-norm relative error (\%) in predicting voltage magnitudes for representative Topology 2 and Topology 7 under the two considered cases. As can be seen, most errors are concentrated around low percentages, with mean values of 0.69\% and 0.47\% for Topology 2 (Case 1) and Topology 7 (Case 2), respectively.  Thus, it indicates the accurate predictions of the proposed NP model.

%across different topologies.

% We further compare our proposed method with the standard Neural process \textcolor{red}{[]} and Deep Sparse GP \textcolor{red}{[]} using metrics including Mean Absolute Error (MAE), Root Mean Square Error (RMSE), L1-norm relative error (\%), and computational cost (in minutes). Further, for the other two methods i.e. standard neural process and Deep Sparse GP, we train separate models for each topology to demonstrate the effectiveness of the proposed approach. The results are shown in Table \ref{tab:diff_errors}. It can be observed that the proposed NP model demonstrates comparable performance for topology 1 and 2 while it performs better for topology 3. The proposed NP model also demonstrates robust performance across different topologies, maintaining relatively low errors and reasonable computational costs. This indicates its adaptability and effectiveness in handling varying system configurations. The standard NP model, while also quite efficient, shows slightly higher errors than proposed NP in Topologies 2 and 3, suggesting that it may not capture the intricacies of all network configurations as effectively as the proposed NP model. The Deep Sparse GP model, although highly accurate in Topology 2, shows significant variability in performance across different topologies, indicating potential limitations in its generalizability. Hence, the proposed NP model stands out for its balanced performance and robustness, making it a strong candidate for applications requiring adaptability to changing network configurations.

% \begin{table}[!ht]
% %\vspace{-0.3cm}
%     \centering
%     \caption{Average errors in estimating voltages (all nodes) on different topologies of IEEE 9-bus using a meta ANP model. }
%   %  \vspace{-0.25cm}
%    \resizebox{0.5\textwidth}{!}{\begin{tabular}{c|lcccc}
%        Topology  & Model & MAE & RMSE & $\%\Vert \Delta \bm{V} \Vert_1$ & Cost (min.)  \\ \hline
%       \multirow{3}{*}{1}   &  Attentive Neural Process & 0.0053 & 0.0068 & 0.49\% & 1.37\\
%        & Neural Process &  0.0051 & 0.0060& 0.47\% &  1.19\\
%       & Deep Sparse GP & 0.0059  & 0.0062& 0.53\% & 2.21\\ \hline
%         \multirow{3}{*}{2} &  Attentive Neural Process & 0.0072  & 0.0099 & 0.65\% & --\\
%         & Neural Process &  0.0079& 0.0100& 0.75\% & -- \\
%           & Deep Sparse GP & 0.0004 &0.0023 & 0.04\% & 0.90\\ \hline
%        \multirow{3}{*}{3}  &  Attentive Neural Process &  0.0072& 0.0092 & 0.67\% & --\\
%        & Neural Process & 0.0085 &  0.0100 & 0.80\% & --\\
%           & Deep Sparse GP& 0.0132 & 0.0169& 1.24\% & 1.38\\ \hline
%     \end{tabular}
%     \label{tab:diff_errors} }
%     %\vspace{-0.25cm}
% \end{table}

\begin{table}[tbp]
%\vspace{-0.3cm}
    \centering
    \caption{Errors in predicting voltage outputs under varying N-1 topologies of IEEE 9-bus using the proposed meta-model NP. }
  \vspace{-0.5em}
   \resizebox{0.5\textwidth}{!}{\begin{tabular}{c|ccccc}
       \textbf{Topology}  & \textbf{Case} & \textbf{MAE} & \textbf{RMSE} & $\%\Vert \Delta \bm{V} \Vert_1$ & \textbf{Cost (min.)}  \\ \hline
      % \multirow{1}{*}{1}   &  Attentive Neural Process & 0.0053 & 0.0068 & 0.49\% & 1.37\\
      \multirow{2}{*}{1}   & Case 1 &  0.0076 & 0.0079& 0.69\% &  2.34\\
      & Case 2 & 0.0036  & 0.0037 & 0.35\% & 3.57\\ 
      \hline
      \multirow{2}{*}{2}   & Case 1 &  0.0073 & 0.0076& 0.67\% &  -\\
      & Case 2 &  0.0120 & 0.0122& 1.11\% & -\\ 
      \hline
      \multirow{2}{*}{3}   & Case 1 &  0.0047 & 0.0050& 0.39\% &  -\\
      & Case 2 &  0.0052 & 0.0056& 0.41\% & -\\ 
      \hline
      \multirow{2}{*}{4}   & Case 1 &  0.0138 & 0.0139& 1.38\% &  -\\
      & Case 2 &  0.0096 & 0.0097 & 0.86\% & -\\ 
      \hline
      \multirow{2}{*}{5}   & Case 1 &  0.0137 & 0.0138& 1.36\% &  -\\
      & Case 2 &  0.0085 & 0.0087& 1.01\% &-\\ 
      \hline
      \multirow{2}{*}{6}   & Case 1 &  0.0120 & 0.0122& 1.25\% &  -\\
      & Case 2 & 0.0081  & 0.0084& 0.70\% & -\\ 
      \hline
      \multirow{2}{*}{7}   & Case 1 &  0.0389 & 0.0391& 3.90\% &  -\\
      & Case 2 &  0.0068 & 0.0080& 0.47\% & -\\ 
      \hline
    \end{tabular}
    \label{tab:diff_errors} }
    \vspace{-0.5cm}
\end{table}

Table \ref{tab:diff_errors} presents the errors in predicting voltage outputs under varying N-1 topologies of the IEEE 9-bus system using the proposed meta-model NP. The table compares the Mean Absolute Error (MAE), Root Mean Square Error (RMSE), and the percentage mean of the $\%L_1$-relative error norm for the two considered cases, i.e. Case 1 (Training a single NP model for all seven N-1 topologies), and Case 2 (Training three separate NP models for three clusters of topologies). For Case 1, the single NP model shows a quite good performance across different topologies, with most errors ranging from 0.39\% to less than 1.4\%, except Topology 7 exhibiting a higher error of 3.90\%. As expected, for Case 2, the clustered NP models generally achieve lower errors, with significant improvements in topologies that exhibit higher errors in Case 1, particularly for topologies with higher complexity, such as Topology 7, where the error is reduced from 3.90\% to 0.47\%. However, the estimation error also shows a slight increase for topologies 2 and 3, but the difference remains insignificant. This reduction in accuracy could be attributed to non-optimal clustering results. As shown in Fig. \ref{fig:topo_Vpc}, we only use a simple relationship plot between PCA and the voltage at bus 4 for clustering. Plotting the relationship between PCA and other nodal voltages may result in different clusters, suggesting the need for a better clustering method.

\vspace{-0.7em}
\subsection{Prediction Accuracy under Severe Contingency}
The prediction accuracies under various line outages are shown in Table \ref{tab:diff_errors}. The highest prediction error was observed to be in Topology 7 (i.e. outage of the line connecting bus 9 to bus 4). 
% The reason behind such behaviour lies in the severity of this contingency. 
The severity of this contingency can be analyzed from Table \ref{tab:line_flow} which shows the loading of each line in the pre-contingency scenarios. 
% The bus 9 receives power from both bus 8 as well as bus 4. It can be observed that the line connecting bus 8 to 9 is heavily loaded and an outage of the line connecting bus 9 to bus 4 further increases the loading of this line. This results in significantly low voltages at bus 9 as can be observed from Fig. \ref{fig:topo_Vdenst}. 
It can also be seen that the voltage under Topology 7 is significantly lower (and even below allowed operational limits) compared to the other topologies. However, a prediction error of just 3.9\% (Case 1) and 0.47\% (Case 2) for this severe contingency shows the effectiveness of the proposed method in predicting voltages under N-1 contingencies. This result also approves our clustering-based approach i.e. the lines with a high contingency ranking should be clustered separately from the other lines whose outage is comparatively less severe.

\begin{table}
    \centering
        \caption{Line flow results in base case.}
    \resizebox{0.45\textwidth}{!}{\begin{tabular}{c|c|c|c}
        Line & Power flow (MVA) & Line  & Power Flow (MVA) \\
            \hline
        Bus 4 to 5 & 30.72 & Bus 7 to 8  & 76.65 \\
        Bus 5 to 6  & 60.96 & Bus 8 to 9  & 87.02\\
        Bus 6 to 7  & 24.38 & Bus 9 to 4 & 56.13 \\
        \hline 
    \end{tabular}
    \label{tab:line_flow} }
    \vspace{-2mm}
\end{table}

\noindent \textbf{Computational Cost: } Regarding computational cost, it is important to note that we only train one proposed meta-model NP under Case 1. Training this single NP model is computationally efficient, requiring just 2.34 minutes. In contrast, training three clustered NP models increases the time to 3.57 minutes. However, this approach offers substantial gains in predictive accuracy, demonstrating the effectiveness of the proposed method and the necessity of topology grouping for handling dynamic network configurations. Note that the reported computational costs are based on training conducted within Google Colab using a single CPU.
%we have to train seven separate conventional models. 
%The proposed NP model, while slightly more computationally expensive than the standard NP model, provides a good balance between accuracy and computational efficiency. In contrast, the Deep Sparse GP model incurs the highest computational cost, with a total training time of 3.49 minutes for the three topologies. 

\vspace{-0.7em}

\subsection{Performance on IEEE 118-bus system}
\begin{figure}[t]
    \centering
    \includegraphics[scale=0.3]{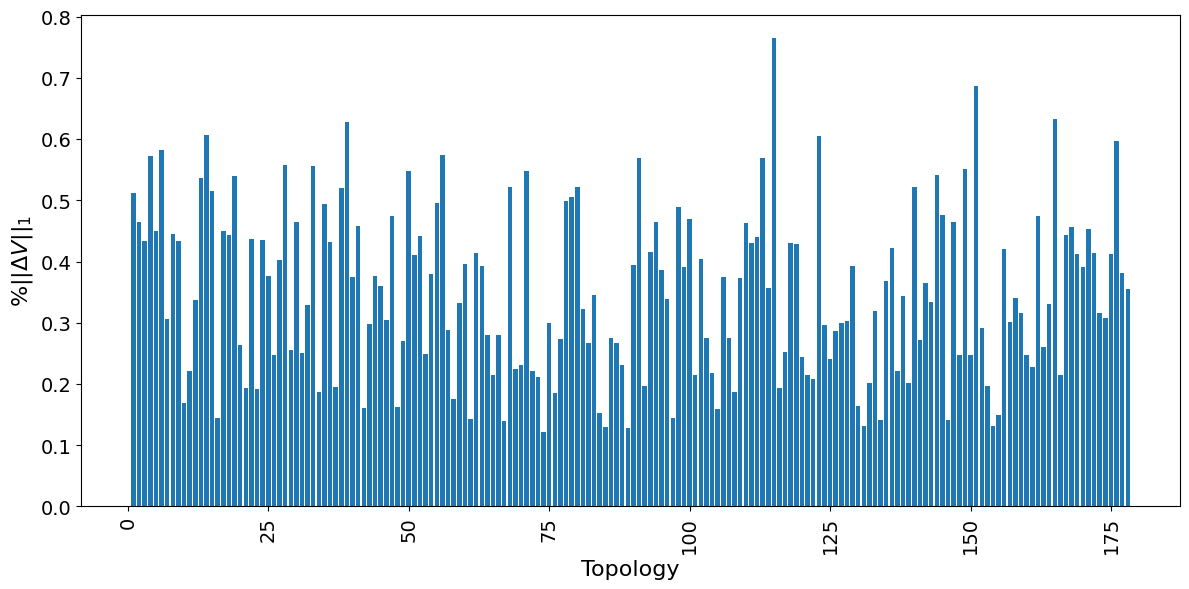}
    \vspace{-8mm}
    \caption{Mean of $L_1$-relative errors (\%) in predicting voltages on IEEE 118-bus under 178 different topologies.}
    \vspace{-4mm}
    \label{fig:118bus}
\end{figure}

The proposed method was further validated on the IEEE 118-Bus test system \cite{118_bus} consisting of 54 generators and 186 transmission lines. The load profiles follow a normal distribution (mean: 0.9 p.u., std: 0.05 p.u.) with a 0.5 correlation between bus demands. Generators at bus 10, 25, 27, 61, 62, and 100 were converted to solar PV farms resulting in approximately 40\% PV penetration with respect to peak demand. The PV output was assumed to be following \textit{Beta} distribution with distribution parameters 2.06 and 2.5 as given in \cite{xu2017power}.
Fig. \ref{fig:118bus} presents bar plots of the mean $L_1$-relative errors (\%) in predicting voltages on the 118-bus system across 178 different topologies. We trained only one model for this network i.e. no clustering was performed. The errors range from a minimum of 0.14\% to a maximum of approximately 0.77\%. This demonstrates that a single trained NP model can also perform well even on a larger-scale network. 
% Also, we observe higher accuracy in the IEEE 118-bus system compared to the IEEE 9-bus system. This could be because a single line outage in a large-scale grid may have less impact than in a small-scale grid.
Note: the total training time with 1000 epochs, as reported by Google Colab using a single GPU T4, is 4.64 minutes.

\vspace{-0.7em}
\section{Conclusion}
This paper proposed a novel meta-model employing the  Neural Process for topology-aware probabilistic power flow. 
%By leveraging attention mechanisms, the NP effectively captures complex relationships within the data, demonstrating enhanced accuracy over standard neural processes  at a marginally higher computational cost. The proposed metamodel also outperforms deep-sparse Gaussian processes  in terms of both computational efficiency and predictive accuracy. In particular, 
Unlike the conventional methods such as GP- or NN-based PPF that require separate training for multiple topologies, the proposed method can be applied by training a single meta-model NP or some clustered NPs, significantly reducing training complexity and computational burden. Our results highlight the potential of NP-based meta-models in advancing efficient and accurate probabilistic power flow for dynamic network configurations. In the future, we will consider advanced NPs such as Attentive NPs and Convolutional NPs for enhancing the performance and expand the proposed approach to the N-2 contingencies.
 %\section*{Acknowledgment}

\vspace{-2mm}
\bibliographystyle{IEEEtran}
\bibliography{Ref}

\begin{thebibliography}{10}
\providecommand{\url}[1]{#1}
\csname url@samestyle\endcsname
\providecommand{\newblock}{\relax}
\providecommand{\bibinfo}[2]{#2}
\providecommand{\BIBentrySTDinterwordspacing}{\spaceskip=0pt\relax}
\providecommand{\BIBentryALTinterwordstretchfactor}{4}
\providecommand{\BIBentryALTinterwordspacing}{\spaceskip=\fontdimen2\font plus
\BIBentryALTinterwordstretchfactor\fontdimen3\font minus
  \fontdimen4\font\relax}
\providecommand{\BIBforeignlanguage}[2]{{%
\expandafter\ifx\csname l@#1\endcsname\relax
\typeout{** WARNING: IEEEtran.bst: No hyphenation pattern has been}%
\typeout{** loaded for the language `#1'. Using the pattern for}%
\typeout{** the default language instead.}%
\else
\language=\csname l@#1\endcsname
\fi
#2}}
\providecommand{\BIBdecl}{\relax}
\BIBdecl

\bibitem{221262}
N.~Hatziargyriou, T.~Karakatsanis, and M.~Papadopoulos, ``Probabilistic load
  flow in distribution systems containing dispersed wind power generation,''
  \emph{IEEE Trans. on Power Systems}, vol.~8, no.~1, 1993.

\bibitem{9516893}
X.~Lin, T.~Shu, J.~Tang, F.~Ponci, A.~Monti, and W.~Li, ``Application of joint
  raw moments-based probabilistic power flow analysis for hybrid ac/vsc-mtdc
  power systems,'' \emph{IEEE Trans. on Power Systems}, vol.~37, no.~2, pp.
  1399--1412, 2022.

\bibitem{8447243}
G.~E. Constante-Flores and M.~S. Illindala, ``Data-driven probabilistic power
  flow analysis for a distribution system with renewable energy sources using
  monte carlo simulation,'' \emph{IEEE Trans. on Industry Applications},
  vol.~55, no.~1, pp. 174--181, 2019.

\bibitem{9870032}
S.~Ly, P.~Pareek, and H.~D. Nguyen, ``Scalable probabilistic optimal power flow
  for high renewables using lite polynomial chaos expansion,'' \emph{IEEE
  Systems Journal}, vol.~17, no.~2, pp. 2282--2293, 2023.

\bibitem{10252798}
S.~Ly, K.~Chauhan, G.~H. Beng, and H.~D. Nguyen, ``A novel quantile lite-pce
  for probabilistic risk assessment of power system cascading outage for n-1-1
  contingency analysis,'' in \emph{2023 IEEE Power \& Energy Society General
  Meeting (PESGM)}, 2023, pp. 1--5.

\bibitem{pareek2021non}
P.~Pareek, C.~Wang, and H.~D. Nguyen, ``Non-parametric probabilistic load flow
  using gaussian process learning,'' \emph{Physica D: Nonlinear Phenomena},
  vol. 424, p. 132941, 2021.

\bibitem{optimalcontrol}
P.~Pareek, W.~Yu, and H.~D. Nguyen, ``Optimal steady-state voltage control
  using gaussian process learning,'' \emph{IEEE Trans. on Industrial
  Informatics}, vol.~17, no.~10, pp. 7017--7027, 2020.

\bibitem{9619978}
N.~Xia, J.~Deng, T.~Zheng, H.~Zhang, J.~Wang, S.~Peng, and L.~Cheng, ``Fuzzy
  logic based network reconfiguration strategy during power system
  restoration,'' \emph{IEEE Systems Journal}, vol.~16, no.~3, 2022.

\bibitem{gao2023physics}
M.~Gao, J.~Yu, Z.~Yang, and J.~Zhao, ``Physics embedded graph convolution
  neural network for power flow calculation considering uncertain injections
  and topology,'' \emph{IEEE Trans. on neural networks and learning systems},
  2023.

\bibitem{xiang2020probabilistic}
M.~Xiang, J.~Yu, Z.~Yang, Y.~Yang, H.~Yu, and H.~He, ``Probabilistic power flow
  with topology changes based on deep neural network,'' \emph{IJEEPS}, vol.
  117, 2020.

\bibitem{du2019achieving}
Y.~Du, F.~Li, J.~Li, and T.~Zheng, ``Achieving 100x acceleration for n-1
  contingency screening with uncertain scenarios using deep convolutional
  neural network,'' \emph{IEEE Trans. on Power Systems}, vol.~34, 2019.

\bibitem{garneloNP}
M.~Garnelo, J.~Schwarz, D.~Rosenbaum, F.~Viola, D.~Rezende, S.~Eslami, and
  Y.~Teh, ``Neural processes,'' in \emph{ICML Workshop on Theoretical
  Foundations and Applications of Deep Generative Models}, 2018.

\bibitem{brunzema2024neural}
P.~Brunzema, P.~Kruse, and S.~Trimpe, ``Neural processes with event triggers
  for fast adaptation to changes,'' in \emph{6th Annual Learning for Dynamics
  \& Control Conference}.\hskip 1em plus 0.5em minus 0.4em\relax PMLR, 2024,
  pp. 1619--1632.

\bibitem{jha2022neural}
S.~Jha, D.~Gong, X.~Wang, R.~E. Turner, and L.~Yao, ``The neural process
  family: Survey, applications and perspectives,'' \emph{arXiv preprint
  arXiv:2209.00517}, 2022.

\bibitem{HEIDARI2022107592}
H.~Heidari, M.~{Tarafdar Hagh}, and P.~Salehpoor, ``Accurate, simultaneous and
  real-time screening of n-1, n-k, and n-1-1 contingencies,'' \emph{JEPES},
  vol. 136, p. 107592, 2022.

\bibitem{nguyen2018constructing}
H.~D. Nguyen, K.~Dvijotham, and K.~Turitsyn, ``Constructing convex inner
  approximations of steady-state security regions,'' \emph{IEEE Transactions on
  Power Systems}, vol.~34, no.~1, pp. 257--267, 2018.

\bibitem{chow1982time}
J.~H. Chow, \emph{Time-scale modeling of dynamic networks with applications to
  power systems}.\hskip 1em plus 0.5em minus 0.4em\relax Springer, 1982.

\bibitem{118_bus}
\BIBentryALTinterwordspacing
R.~Christie, ``118 bus power flow test case,'' May 1993, [Accessed 28-01-2025].
  [Online]. Available: \url{https://labs.ece.uw.edu
  /pstca/pf118/pg\_tca118bus.htm}
\BIBentrySTDinterwordspacing

\bibitem{zimmerman2010matpower}
R.~D. Zimmerman, C.~E. Murillo-S{\'a}nchez, and R.~J. Thomas, ``Matpower:
  Steady-state operations, planning, and analysis tools for power systems
  research and education,'' \emph{IEEE Trans. on power systems}, vol.~26,
  no.~1, pp. 12--19, 2010.

\bibitem{xu2017power}
X.~Xu, Z.~Yan, M.~Shahidehpour, H.~Wang, and S.~Chen, ``Power system voltage
  stability evaluation considering renewable energy with correlated
  variabilities,'' \emph{IEEE Transactions on Power Systems}, vol.~33, no.~3,
  pp. 3236--3245, 2017.

\end{thebibliography}

\end{document}